\documentclass[12pt]{iopart}
\usepackage{graphicx}  
\usepackage{subfigure}

\newcounter{shimeqsno}
\def\av#1{\langle #1 \rangle}
\def\bsigma{\mbox{\boldmath $\sigma$}}
\def\bxi{\mbox{\boldmath $\xi$}}
\def\erf{\mbox{\rm erf}}
\def\e{\mbox{\rm e}}
\def\sign{\mbox{\rm sign}}
\def\Tr{\mbox{\rm Tr}}

\newcommand{\si}{\sigma}

\begin{document}
\jl{1}
\title
{Thermodynamic properties of extremely diluted symmetric $Q$-Ising neural
networks}
\author{D Boll\'{e}\dag\footnote[3]{Also at Interdisciplinair 
Centrum voor Neurale Netwerken, K.U.Leuven, Belgium\\e-mail:
desire.bolle@fys.kuleuven.ac.be, domenico.carlucci@fys.kuleuven.ac.be,
 gmshim@cnu.ac.kr },  
D M Carlucci\dag \,\,and G M Shim\ddag}
\address{\dag\ Instituut voor Theoretische Fysica,
K.U. Leuven, B-3001 Leuven, Belgium}
\address{\ddag\ Department of Physics, Chungnam National University,
Yuseong, Taejon 305-764, R.O. Korea}

\begin{abstract}
Using the replica-symmetric mean-field theory approach the thermodynamic
and retrieval properties of extremely diluted {\it symmetric} $Q$-Ising 
neural networks are studied. In particular, capacity-gain parameter
and capacity-temperature phase diagrams are derived for $Q=3,\,4$ and
$Q=\infty$. The zero-temperature results are compared with those obtained 
from a study of the dynamics of the model. Furthermore, the de 
Almeida-Thouless line is determined. Where appropriate, the difference with
other $Q$-Ising architectures is outlined. 
\end{abstract}


\maketitle

\section{Introduction}

Recently the dynamics of extremely diluted {\it symmetric} $Q$-Ising
neural networks has been solved completely \cite{bjs99}. In spite of the
extremely diluted architecture, precisely the symmetry causes feedback
correlations from the second time step onwards, in contrast with an
asymmetric architecture \cite{dgz87,bsvz94}, complicating the dynamics
in a nontrivial way. Based upon the time evolution of the  distribution
of the local field a recursive scheme has been developed in order to
calculate the evolution of the relevant order parameters of the network at
zero temperature. Furthermore, by requiring that the local field becomes
time-independent implying that some correlations are neglected,
fixed-point equations are obtained  and the capacity-gain parameter
phase diagrams for the $Q=2,\,3$ models have been discussed shortly.

For the case of $Q=2$ it turns out that the resulting fixed-point
equations from this dynamic approach are the same as those derived from
a thermodynamic replica-symmetric mean-field theory treatment given in
ref.~\cite{ws91}. This gives us some insight concerning a possible
relation between replica symmetry and the structure of the noise. Of
course, it would be interesting to know whether this stays valid for
$Q>2$. Besides, a detailed discussion of the phase diagram for these
extremely diluted symmetric models is interesting on its own. However,
for these cases a thermodynamic approach is
not yet available in the literature. The purpose of this work is
precisely to present the results of such an approach. 

Concretely, we consider the extremely diluted symmetric $Q$-Ising
neural network with arbitrary gain parameter. Using the
replica-symmetric mean-field approximation we investigate both its
thermodynamic and retrieval properties at zero and non-zero
temperatures. Explicit results are presented for $Q=3,\,4$, and
$Q=\infty$. 

The rest of this paper is organized as follows. In section 2 the model
is introduced from a dynamical point of view. Section 3 presents the
replica-symmetric mean-field approximation and obtains the relevant
fixed-point equations for general $Q$. In section 4 these equations are
studied in detail for arbitrary temperatures and $Q=3$ (section 4.1), $Q=4$
(section 4.2) and $Q=\infty$ (section 4.3). In particular, the storage
capacity as a function of the gain parameter and capacity-temperature
phase diagrams are obtained. The specific thermodynamic properties are
discussed. The results turn out to be significantly different for odd and
even $Q$. They are compared with those for other architectures of these 
$Q$-Ising
models. Section 6 presents the concluding remarks. Finally, the appendix
contains the specific fixed-point equations for the different values of $Q$
treated in the paper.

\section{The model}

Consider a network of $N$ neurons which can take values $\sigma_i$ in the 
set of equidistant states
\begin{equation}
      {\cal S}_Q = \{s_k=-1+2(k-1)/(Q-1),\;k=1, \ldots,Q\}.
      \label{eq:r6}
\end{equation}
In this network we want to store $p$ patterns
$\{\bxi^\mu,\,\mu=1,\ldots,p\}$ that are supposed to be
independent and identically distributed random variables (i.i.d.r.v.) with 
zero mean and variance $A$. The latter is a measure for the activity of
the patterns.

Given a configuration $\bsigma = (\sigma_1,\ldots,\sigma_N)$, the local
field $h_i$ of neuron $i$ is
\begin{equation}
         h_i(\bsigma) = \sum_{j\ne i} J_{ij}\sigma_j\,,
         \label{eq:r1}
\end{equation}
with $J_{ij}$ the synaptic couplings between neurons $i$ and $j$.

The network is taken to be extremely diluted but symmetric meaning that
the couplings are chosen as follows. Let $\{c_{ij}=0,1\}, i,j= 1,
\ldots, N $ be i.i.d.r.v. with distribution
$\mbox{Pr}\{c_{ij}=x\}=(1-c/N)\delta_{x,0} + (c/N) \delta_{x,1}$ and
satisfying $c_{ij}=c_{ji},\,\,\,c_{ii}=0 $, then
\begin{equation}
        \label{eq:r2}
  J_{ij}=\frac{c_{ij}}{Ac} \sum_{\mu=1}^p \xi_i^\mu \xi_j^\mu
        \quad \mbox{for} \quad i \not=j            \,.
\end{equation}
We remark that compared with the asymmetrically diluted model \cite{bsvz94}
the architecture is still a local Cayley-tree  but no longer directed.  
In the limit $N \rightarrow \infty$ the probability that the set of 
connections, $T_i=\{j=1, \ldots, N |c_{ij}=1\}$, giving information to the
site $i$, is equal to a certain number $k$ remains a Poisson distribution 
with mean $c=E[|T_i|]$.
Thereby it is assumed that $ c \ll \log N$. In order to get an infinite
average connectivity allowing to store infinitely many patterns $p$, one also
takes then the limit $c \rightarrow \infty$ and defines the capacity $\alpha$ 
by $p= \alpha c$.
However, although for the asymmetric architecture, at any given time step
$t$ all spins are uncorrelated and hence no feedback is present, for the
symmetric architecture this is no longer the case, causing feedback
from $t \geq 2$ onwards \cite{ws91,pz90}. Indeed, if the coupling $J_{ij}$ is
non-zero then also $J_{ji}$ is non-zero and therefore the state of neuron $i$
at time $t$ depends on its state at time $t-2,t-4, \ldots$. This is not
the case for asymmetric dilution since the probability to have a
non-zero $J_{ji}$ given a non-zero $J_{ij}$, vanishes.

The neurons are updated asynchronously according to the transition
probability
\begin{equation}
   \Pr(\sigma_i'=s_k|\bsigma) = \frac{\exp[-\beta\epsilon_i (
       s_k|h_i(\bsigma))] }{\sum_{l=1}^Q \exp[-\beta\epsilon_i(
       s_l|h_i(\bsigma))] }\,.
       \label{eq:r3}
\end{equation}
Here the inverse temperature $\beta=T^{-1}$ measures the noise level, and
the energy potential $\epsilon_i(s|h)$ is taken to be \cite{r90}
\begin{equation}
     \epsilon_i(s|h) = -hs+bs^2  \quad b>0
     \label{eq:r4}
\end{equation}
At zero temperature, $\sigma_i'$ takes
the value $s_{k}$ leading to the minimum of the energy potential. This is
equivalent to using an input-output relation
\begin{eqnarray}
       \sigma_i' &=& g_b[h_i(\bsigma)] \nonumber \\
            g_b(x) &=&\sum_{k=1}^Qs_{k} [
             \theta(b(s_{k+1}+s_{k})-x)
            - \theta(b(s_{k}+s_{k-1})-x)]
    \label{eq:r5}
\end{eqnarray}
with $s_{0} = -\infty$ and $s_{Q+1} = \infty$.
For finite $Q$ this input-output relation has a steplike shape and
the parameter $b$ controls the steepness of the steps.
For $Q=\infty$ the input-output function (\ref{eq:r5}) becomes the
piecewise linear function
\begin{equation}
     g_b(x) = \left\{ \begin{array}{ll}
                        \sign(x) &\mbox{if $|x|>2b$} \\
                       \frac{x}{2b} &\mbox{otherwise.}
                     \end{array} \right.
     \label{eq:r7}
\end{equation}
The slope of the linear part is given by $(2b)^{-1}$. In general, as
$b$ goes to zero the input-output relation reduces to that of the
Ising-type network independent of $Q$.

In what follows we present a detailed study of the properties
of these symmetrically diluted networks as a function of $T$ and $b$.

\section{Replica-symmetric mean-field theory}

The longtime behaviour of the network under consideration is governed by
the Hamiltonian
\begin{equation}
     H \,=\,
       -\frac{1}{2}\sum_{i \neq j} J_{ij} \sigma_i\sigma_j 
        \,+\, b\sum_i \sigma_i^2
     \label{ham}	
\end{equation}

In order to calculate the free energy we use the standard replica method as
applied to dilute spin-glass models \cite{ws91,cn92,vb85,bcs76}. Starting from
the replicated partition function averaged over the connectivity  and
the non-condensed patterns we arrive at
\begin{eqnarray}
\fl && \langle {\cal Z}^n   \rangle_{\rm c}
   \,=\,
  \prod_{\mu,\gamma} \left[\int  d m_{\gamma}^{\mu} \right] 
   \prod_{\gamma, \gamma'}\left[ \int  d q_{\gamma \gamma'} \right]
   \prod_{\gamma} \left[\int   d {\tilde q}_{\gamma} \right]  
                  \,\exp\left[ - N \beta f \right]
  \label{part}	 
         \\
\fl && f\,=\,
    \frac{A}{2}\sum_{\mu,\gamma} \left( m_{\gamma}^{\mu} \right)^2 
   +
    \frac{\beta \alpha}{2}\sum_{\gamma<\gamma'} q_{\gamma\gamma'}^2 
   + 
    \frac{\beta \alpha}{2}\sum_{\gamma} {\tilde q}_{\gamma}^2 
   -
   \frac{1}{\beta}
      \left\langle\kern-0.3em\left\langle 
          \ln_{ \{\si^{\alpha}\}}{\rm Tr}
          \exp\left[-\beta \widetilde{H}\right]
      \right\rangle\kern-0.3em\right\rangle_{\{   \xi\}} 
  \label{free1}
        \\
\fl  &&\widetilde{H}= - \sum_{\mu,\gamma} 
         m_{\gamma}^{\mu}\xi^{\mu}\sigma_{\gamma} 
         \,-\beta \alpha 
      \sum_{\gamma<\gamma'} q_{\gamma\gamma'}\sigma_{\gamma}
                                    \sigma_{\gamma'} 
     +\sum_{\gamma}
          \left(b - \frac{\alpha\beta}{2}{\tilde q}_{\gamma}\right)
     \sigma_{\gamma}^2
   \label{free2}    
\end{eqnarray}
where the $m_{\gamma}^{\mu}, q_{\gamma \gamma'}, {\tilde q}_{\gamma}$ are
the usual order parameters defined by 
\begin{eqnarray} 
   m_{\gamma}^{\mu} = \frac{1}{N} 
               \sum_i \xi_i^{\mu} \left\langle\sigma_{\gamma,i} 
	       \right\rangle  \quad \gamma=1, \ldots, n \\
   q_{\gamma \gamma'}= \frac{1}{N}
   	       \sum_i \left\langle\sigma_{\gamma,i}\right\rangle 
	       \left\langle\sigma_{\gamma',i} \right\rangle 
	        \quad \gamma \neq \gamma'=1, \ldots, n \\
   {\tilde q}_{\gamma} = \frac{1}{N}
           \sum_i \left\langle\sigma_{\gamma,i}^2\right\rangle 
	        \quad \gamma=1, \ldots, n 
   	\label{order}       
\end{eqnarray}
with $n$ the number of replicas and the sum over $\mu$ running over the
number of condensed patterns
$s$. Assuming replica symmetry, i.e., $m_{\gamma}^{\mu}= m_{\mu},
 q_{\gamma \gamma'}=q, {\tilde q}_{\gamma}=\tilde q$ the free energy
becomes 
\begin{eqnarray}
 \fl f(\beta) \,=\, \lim_{n \to 0} \frac{f}{n} \,=\,
   \frac{A}{2}\sum_{\mu} \left( m_{\mu} \right)^2 
        + \frac{ \alpha}{4 \beta} \chi^2 
        + \frac{\alpha}{2} q \chi 
             \nonumber     \\
\fl \hspace*{2.7cm} \,-\,
  \frac{1}{\beta}
  \left\langle\kern-0.3em\left\langle 
      \int {\cal D} z 
      \ln_{ \{ \si \}   }{\rm Tr}
         \exp\left[\beta \sigma
	    \left(\sum_{\mu} m_{\mu} \xi^{\mu}+ \sqrt{\alpha q} \, z 
                - {\tilde b}\sigma \right)
             \right]
          \right\rangle\kern-0.3em\right\rangle_{\{\xi\}} 
	\label{freenew}
\end{eqnarray}
with $ {\tilde b}\equiv b- \frac{\alpha}{2}\chi$ the effective gain
parameter, 
$\chi\equiv\beta({\tilde q}-q)$ the susceptibility and ${\cal D} z= 
dz\,(2\pi)^{-1/2} \exp(-z^2/2)$ the Gaussian measure. 
We remark that the effective gain parameter $\tilde b$ can be negative,
implying that the input-output function reduces to that of $2$-Ising-type
neurons, i.e., $g_{\tilde{b}}(h) = \mbox{sign}(h)$. Furthermore, we note
that the free energy (\ref{freenew}) can be obtained as the {\it formal} 
expansion to second order in $\chi$ of the free energy 
for the fully connected model derived in \cite{brs94}.

The phase structure of the network is determined by the
solution of the fixed-point equations for the order parameters
\begin{eqnarray}
    m_\mu &=& \frac1A \left \langle \! \left \langle
             \int Dz\, \xi^\mu \av{\sigma(z)}
             \right \rangle \! \right \rangle
              \label{eq:s6} \\
    q &=& \left \langle \! \left \langle
               \int Dz\, \av{\sigma(z)}^2
               \right \rangle \! \right \rangle
               \label{eq:s7} \\
    \chi &=& \frac1{\sqrt{\alpha q}} \left \langle \! \left \langle
        \int Dz\, z\,\av{\sigma(z)} \right \rangle \! \right \rangle 
                \label{eq:s8}
\end{eqnarray}
which extremize $-\beta f (\beta)$. Here
\begin{equation}
      \av{\sigma(z)} = \frac{\Tr_{\sigma} \sigma \exp[\,
           \beta\sigma\,(\sum_\mu m_\mu \xi^\mu+\sqrt{\alpha q}\,z
           -\tilde{b}\sigma)]}
           { \Tr_s \exp[\, \beta s\,(\sum_\mu
           m_\mu \xi^\mu + \sqrt{\alpha q}\,z-\tilde{b} s)]}\,.
           \label{eq:s9}
\end{equation}
In the following section we discuss these equations for $Q=3,4$ and
$Q=\infty$ models.

\section{ Thermodynamic and retrieval properties}

\subsection{$Q=3$}

In the rest of this work we are mainly interested in the Mattis retrieval 
state so that we consider one condensed pattern, say the first one.
Then we can write $m_\mu=m\delta_{\mu 1}$ and, furthermore, we 
leave out the index $1$ in the sequel.

Let us consider a three-state network with uniformly distributed 
patterns taking the values $0,\pm1$ such that $A=2/3$. For the Mattis
retrieval state the average (\ref{eq:s9}) is given by
\begin{equation}
    \langle \sigma(z) \rangle = 
       \frac{\sinh[\beta(m\xi+\sqrt{\alpha q}\,z)] }
                        {\frac12\exp(\beta \tilde{b})
          +\cosh[\beta(m\xi+\sqrt{\alpha q}\,z)] }
   \,, \label{ta3}
\end{equation}
which reads at zero temperature
\begin{equation}
    \langle \sigma(z) \rangle = 
           g_{\tilde{b}}( m\xi+\sqrt{\alpha q}\,z )  \,.
\end{equation}

The fixed-point equations (\ref{eq:s6})-(\ref{eq:s8}) can then easily be
written down and the integration can be performed explicitly at zero
temperature. For completeness we present them in the Appendix. 

First we look at the network at zero temperature.
For $\tilde{b} \leq 0$ these fixed-point equations can be further reduced
by taking the limit $\tilde{b} \rightarrow 0$ and introducing the
variable $x=m/\sqrt{2 \alpha q}$. One arrives at
\begin{equation}
      \sqrt{2 \alpha} = \frac{\erf(x)}{x}
\end{equation}
with, in view of the definition of $\tilde{b}=b- \frac{\alpha}{2}\chi$
\begin{equation}
        b \leq b_0= \sqrt{\frac{\alpha}{18\pi}}\left[\,
                      2\exp(-x^2)+1 \right] \, .
\end{equation}
This learns us that the retrieval state vanishes continuously at 
$\alpha_c = 2 /{\pi}$ for $b \le \ b_0= 1 / {\pi}$. 

\begin{figure}[t]
   \vspace{1.5cm}
       \centering
      \includegraphics[height=0.5\textwidth,
             angle=0]{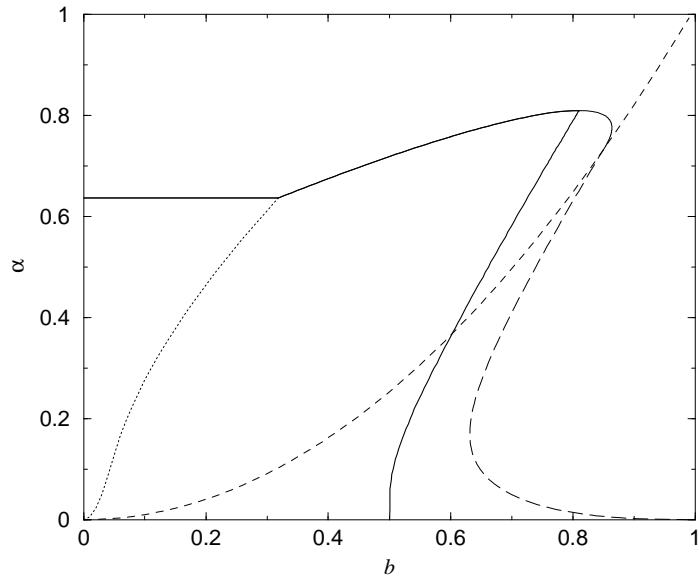}
       \caption{$Q=3$ $\alpha -b$ phase diagram for uniform patterns at $T=0$.
      The (thin) full and long-dashed lines denote the boundary of the
      retrieval region corresponding to respectively a continuous and
      discontinuous appearance of the solution. The dotted line
      separates the  $2$-Ising-like retrieval region. The short-dashed line 
      indicates the discontinuous appearance of the spin-glass state.  
      The thick full line represents the thermodynamic transition for the
      retrieval state.}
       \label{fig:q3t0retr}
\end{figure}

The phase boundary of the retrieval state for $\tilde{b}>0$ 
can be obtained by numerically solving the fixed-point equations
(\ref{q3T0spe1})-(\ref{q3T0spe3}). The results
are shown in the capacity-gain parameter phase diagram presented in
figure \ref{fig:q3t0retr} as the full and the long-dashed curve. The full
curve denotes a continuous appearance of the retrieval state, the 
long-dashed curve a discontinuous one.
Furthermore, the dotted curve separates the $2$-Ising-like region (which
is to the left) where $\tilde{b} \leq 0$. 

Finally, the short-dashed line in this diagram indicates the boundary
above which the spin-glass solution exists. 
The spin glass is characterized by taking $m=0$ and $q\ne0$ in 
(\ref{q3T0spe1})-(\ref{q3T0spe3}). 
The resulting fixed-point equations can be combined into a single
equation in the variable $y=\tilde{b}/\sqrt{2 \alpha q}$ 
\begin{equation}
    \frac{b}{\sqrt{2\alpha}} = 
    y\sqrt{1-\erf(y)}\, + \frac{  \exp(-y^2)  }{
                                2\sqrt{[\pi(1-\erf(y))]} }
    \,.
    \label{spinnew}
\end{equation}
Since the right-hand side of the above equation is bounded, 
a spin-glass solution is possible when $\alpha \ge cb^2$ with $c\approx
1.0134$. It vanishes discontinuously at the boundary.

In order to find the thermodynamic transition line we have to determine 
which state -- retrieval, spin-glass or paramagnetic -- leads to the
lowest free energy. At zero temperature the free energy is given by 
\begin{equation}
      f = -\frac{A}{2} m^2-\frac{\alpha}{2}q\chi+\tilde{b}q
     \label{feT0}
\end{equation}
as long as $\chi$ is finite. This is certainly the case when $q$ is nonzero.
Since also $q=0$ for the paramagnetic state, however, it is necessary to
check whether $\chi$ is finite. The paramagnetic state is determined by the
equation
\begin{equation}
    \tilde{b} = b - \frac{\alpha\beta}{\exp(\beta\tilde{b})+2} \,.
    \label{para}
\end{equation}
This equation has three solutions. In the limit $T\rightarrow 0$, only
the solution that converges to $b$ with $\chi \sim 2 \beta \exp(- \beta
b) \rightarrow 0$  is stable against longitudinal fluctuations as is easily 
seen by expanding the free energy for small q. The free energy of this
stable paramagnetic state is always zero.  
The retrieval state becomes the global minimum of the free energy in the
region bounded by the thick full curve. So, for $\alpha=0$ the retrieval 
state is the minimum for $b \in [0,1/2]$. The latter is true independent 
from the architecture. To the right of the thick full curve the
paramagnetic state is the global minimum. We remark that
this paramagnetic state at $T=0$ is in fact simply a frozen state with
all the spins taking the value zero (and not a phase where all the spins
take all possible values with the same probability).

Finally, the stability of the replica symmetric retrieval solution against
replica-symmetry breaking is determined by the replicon eigenvalue
\cite{mpv78}
\begin{eqnarray}
  \lambda_R \, &=& \,
        1 \,-\,
         \beta^2 \alpha \left\langle\kern-0.2em\left\langle
         \int {\cal D}z\,
    \left(\langle \sigma^2(z) \rangle - \langle \sigma(z)\rangle^2\right)^2   
            \right\rangle\kern-0.2em\right\rangle_{\{\xi\}}  
              \\
          \, &=& \,  1 - 
       \frac{1}{q} \left\langle\kern-0.2em\left\langle
           \int {\cal D}z\, \biggl(    
                \partial_{z} \langle \sigma(z) \rangle
           \biggr)^2   
       \right\rangle\kern-0.2em\right\rangle_{\{\xi\}} 
       \label{replinew} 
\end{eqnarray} 
where $\langle \sigma(z) \rangle$ is given by eq.~(\ref{ta3}). 
In the limit $T\rightarrow0$, $\langle \sigma(z) \rangle$ becomes
a step function so that the argument of the integration over $z$
is proportional to the square of the delta function.
Therefore the replicon eigenvalue for the retrieval solution at $T=0$ is
always negative and, hence, replica symmetry is broken.

For the paramagnetic solution that we have discussed before we find that the
replicon eigenvalue $\lambda_{PM} = 1$ so that this state is also stable
against transverse fluctuations.

At this point a couple of remarks are in order. First, we note that the 
fixed-point equations and, hence, also the $\alpha-b$ phase
diagram are precisely the same as those obtained from an exact
dynamical approach (see \cite{bjs99} eqs.~(23)-(25) and fig.~2) after 
requiring that the evolution of the distribution of the local field becomes 
stationary at a certain time $t$.
This requirement has the consequence that most of the discrete noise
caused by the feedback in the retrieval dynamics is neglected. It would
mean that for these models a replica-symmetric mean-field theory
treatment corresponds to a specific simplification of the structure of
the noise -- only the Gaussian part plus the discrete noise coming from
the last time step  -- induced by the non-condensed patterns. This is
also related to the so called self-consistent signal-to-noise-ratio
analysis introduced in \cite{sf90}. However, more work is needed, also
on other models, to put such a type of conjecture on a firm  basis.
Second, it is interesting to compare the phase diagram
figure \ref{fig:q3t0retr} with the one for the
extremely diluted {\it asymmetric} architecture \cite{bsvz94}. The essential
difference is that the $2$-Ising-like region is much more extended here.
The rest of the diagram has, in fact, a similar shape but it is tilted
because of this Ising-like region towards
greater $b$-values (compare \cite{bsvz94} fig.~1).

\begin{figure}
       \subfigure[]{
             \includegraphics[height=0.40\textwidth,
                               angle=0]{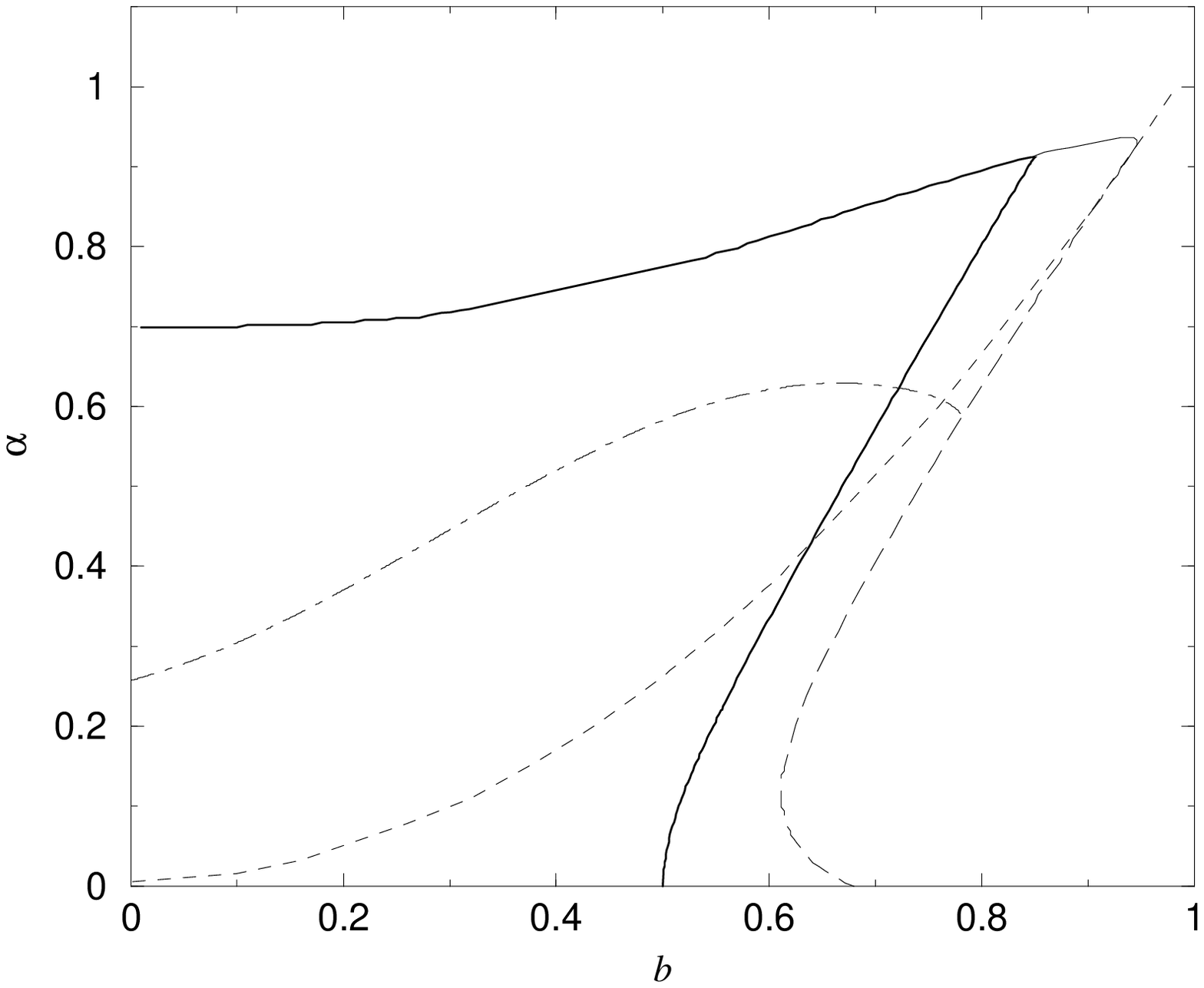}}
     	 \hspace{0.3cm}
	\subfigure[]{
             \includegraphics[height=0.40\textwidth,
                              angle=0]{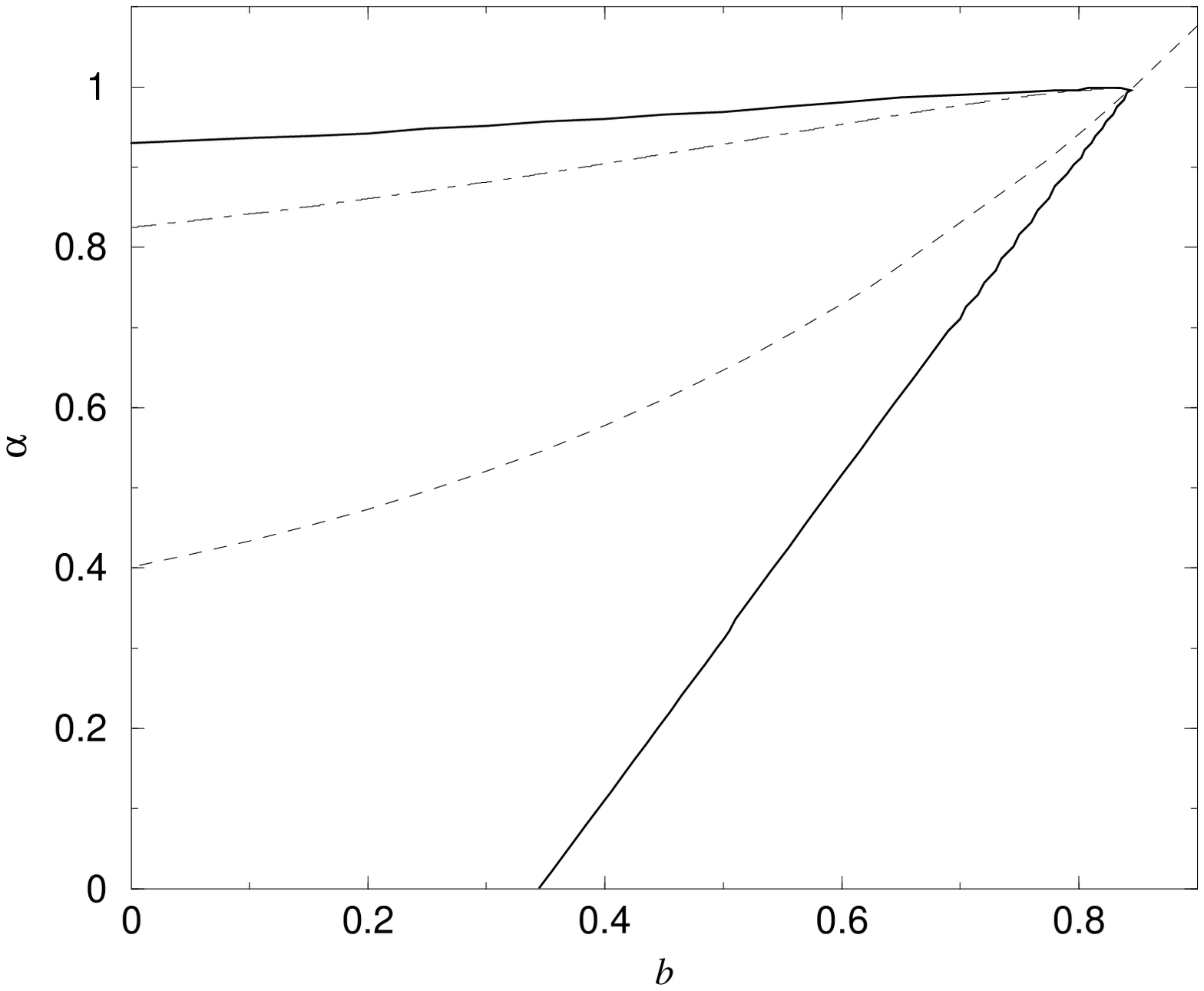}}     
	     \caption{$Q=3$ $\alpha -b$ phase diagram for uniform patterns at 
       $(a)$: $T=0.1$ and $(b)$: $T=0.5$. The meaning of the lines is as 
       in Fig.\ 1. The dashed-dotted curve is the de Almeida-Thouless line
       above which replica symmetry is unstable.} 
       \label{fig:q3tn0b}
\end{figure}

\begin{figure}
   \vspace{1.5cm}    
       \subfigure[]{
             \includegraphics[height=0.4\textwidth,
                               angle=0]{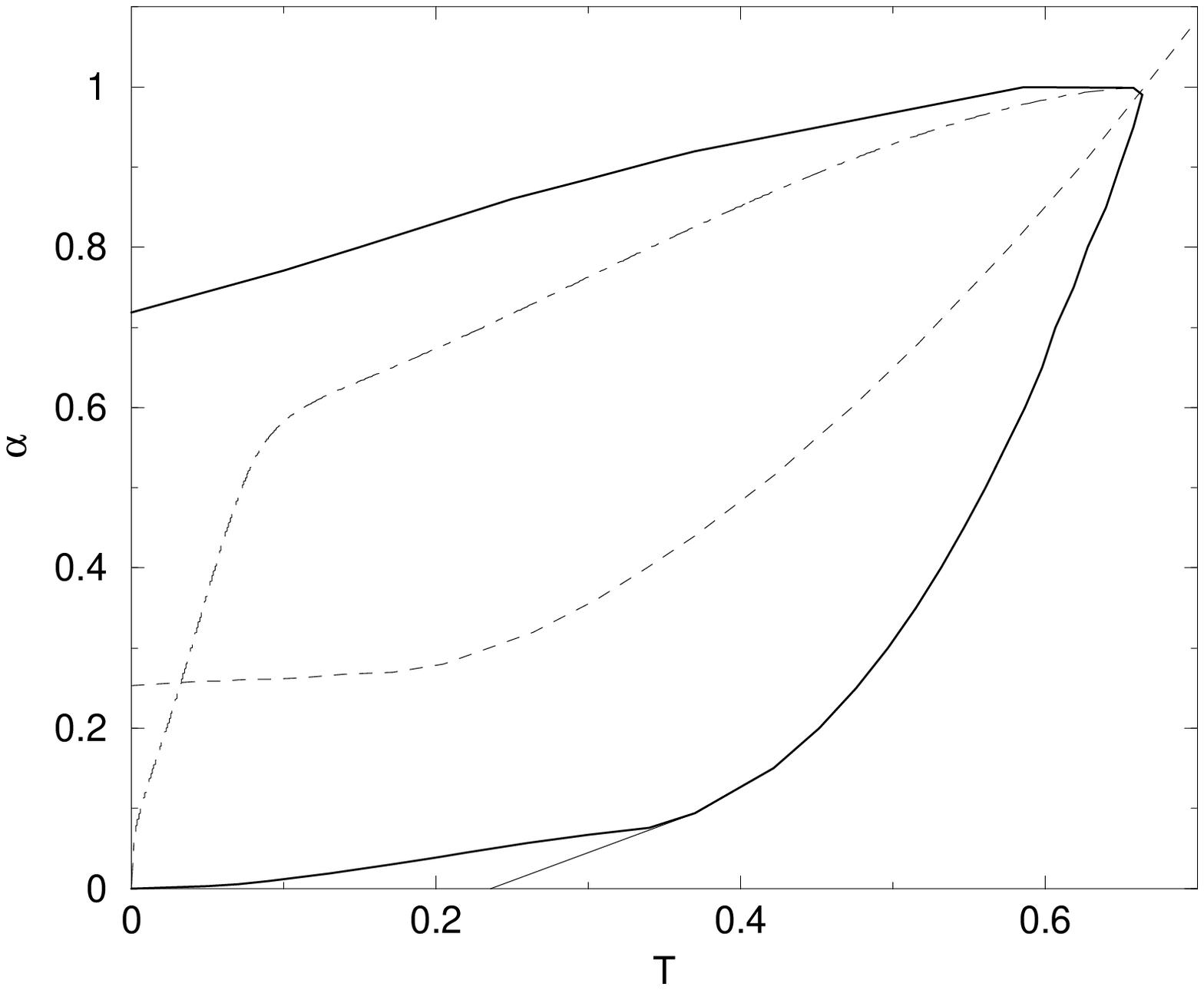}}
     	 \hspace{0.3cm}
	\subfigure[]{
             \includegraphics[height=0.4\textwidth,
                              angle=0]{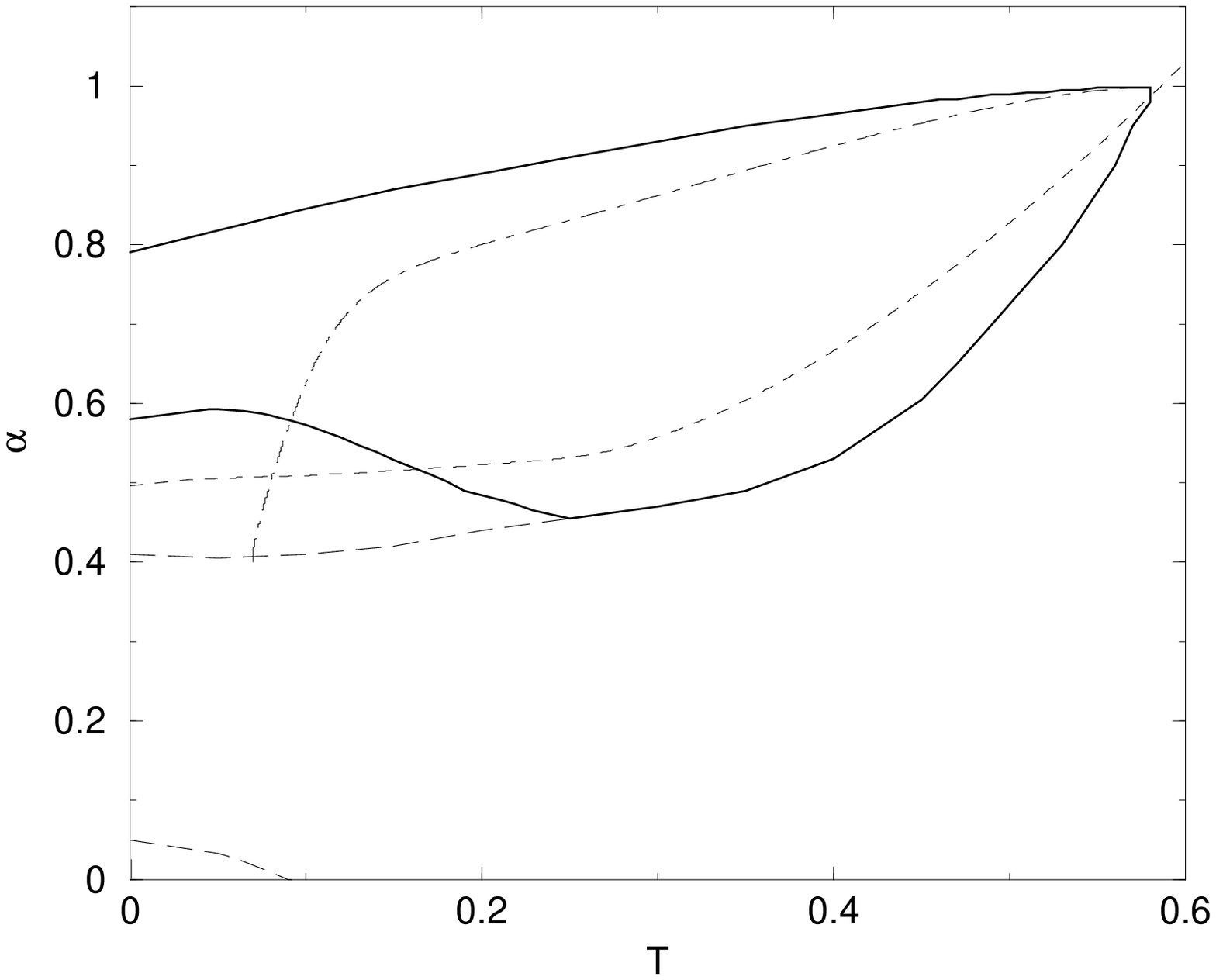}}     
	     \caption{$Q=3$ $\alpha -T$ phase diagrams for uniform
	     patterns  and $(a)$: $b=0.5$ and $(b)$: $b=0.7$,
             The meaning of the lines is as in Figs.\ 1 and 2. } 
       \label{fig:q3tn0T}
\end{figure}

Next we consider the $Q=3$ network at non-zero temperatures. The results
are presented in figures \ref{fig:q3tn0b} and \ref{fig:q3tn0T}. 
From $T=0.37$  onwards the retrieval state appears continuously at
the whole boundary. Furthermore, the retrieval state is the global
minimum of the free energy in a growing region covering the whole
retrieval region of the phase diagram as the  temperature increases to
$T=0.5$. 
For $T=0.6$ the lower branch of the retrieval boundary
ends in zero such that for higher temperatures we apparently have no
retrieval for small values of $\alpha$. For $T=0.7$ and $b=0$, e.g., $\alpha
\geq 0.2$.
Finally, the replica symmetry solution is no longer unstable in the
whole  retrieval
region. Calculating the replicon eigenvalue (\ref{replinew}) leading to
the de Almeida-Thouless (AT) line (dashed-dotted curve in figures 
\ref{fig:q3tn0b} and \ref{fig:q3tn0T}) we find a growing region of 
stability in the phase diagram below this AT-line. 

In order to give a more complete idea of these results we present the 
capacity-temperature phase diagrams in figure \ref{fig:q3tn0T}.
We immediately notice that for higher temperatures we get a higher
critical capacity, in other words the upper branch of the retrieval
boundary shows some strong re-entrant behaviour. This 
is not surprising since this branch lies completely in the region where
replica symmetry is broken, in contrast with the fully-connected
(symmetric) model \cite{brs94}. It is common knowledge, at least for
$Q=2$ \cite{ws91,cn92}, that re-entrance is due to the use of the 
replica-symmetric approximation. It is also conjectured there that a
full replica symmetry breaking solution might be a horizontal line above
this upper branch of the retrieval boundary starting from the high
temperature point. Validating this conjecture by 
numerical simulations or doing a one step replica symmetry breaking 
calculation in order to see whether such results lie indeed closer to
this horizontal line are beyond the scope of the present work.

\subsection{$Q=4$}

The thermodynamic properties of a network consisting
of neurons that are able to take on the zero--state ($\sigma_i=0$)
are significantly different from those of a network in which this
state is forbidden for the neurons. Indeed, for the even $Q$ models a
paramagnetic phase at zero temperature has to be absent precisely
because the spins can not take the value zero and, hence, $q$ cannot be
zero physically. For high temperatures, of course, $q$ can be zero and
such phase does exist. This is similar to the asymmetric diluted 
architecture \cite{bsvz94} and the fully connected architecture 
\cite{brs94}.  As a representative example we consider the $Q=4$ model 
in this section.

According to (\ref{eq:r6}) the neurons, as well as the patterns, can take
on the values $-1,-1/3,+1/3,+1$.  We consider uniformly distributed
patterns so that $A=5/9$.

At zero temperature, the fixed-point equations for a retrieval state
with $\tilde{b} > 0$ are given in the appendix.
For $\tilde{b}< 0$, these equations can be further reduced
by the introduction of the variable $x=m/{\sqrt{2\alpha q}}$ to
\begin{equation}
      \sqrt{2\alpha} = \frac{9}{10x} \biggl[
                       \erf(x)+\frac13\erf(\frac13x) \biggr]
      \label{q4single}
\end{equation}
with, in view of the definition of $\tilde b$
\begin{equation}
    b \leq b_0= \sqrt{\frac{\alpha}{8\pi}}\biggl[
                \exp(-x^2)+\exp(-\frac19x^2) \biggr]
\end{equation}

\begin{figure}
       \centering
             \includegraphics[height=0.5\textwidth,
                              angle=0]{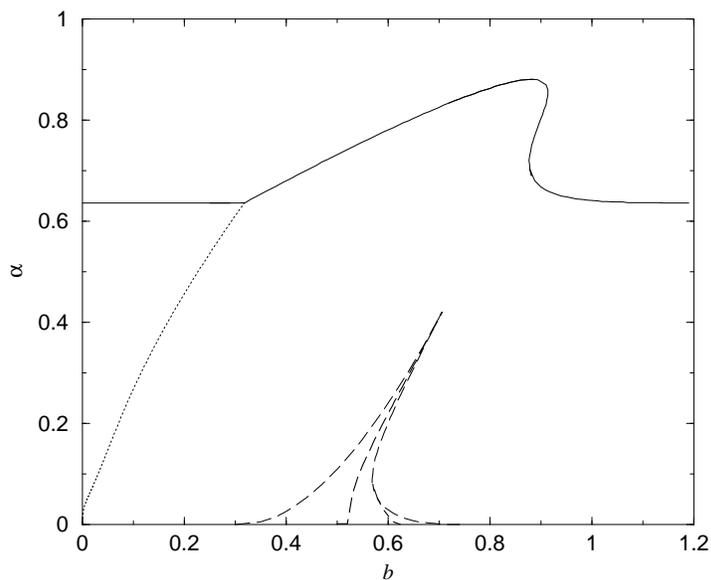}      
	     \caption{$Q=4$ $\alpha -b$ phase diagrams for uniform
	     patterns at  $T=0$. 
       The meaning of the lines is as in Fig.\ 1. The triangular region
       is explained in the text.} 
      \label{fig:q4}
\end{figure}

Exactly as in the $Q=3$ case the retrieval state vanishes continuously
at $\alpha_c=2/{\pi}$ for $b \leq b_0= 1/{\pi}$. 
In contrast with $Q=3$, however, the fixed-point equations have retrieval
solutions for all values of $b$.
When $b\rightarrow\infty$, they can also be reduced
to a single equation similar to eq.~(\ref{q4single}). This means that the
retrieval state again vanishes continuously at $\alpha_c=2/{\pi}$.
For the other values of $b$ the retrieval phase boundary can be obtained
numerically from (\ref{q4mt0})-(\ref{q4ct0}). The result is indicated in 
figure \ref{fig:q4} by a full line expressing the fact that the
retrieval state appears continuously everywhere.

The triangular region in figure \ref{fig:q4} bordered by the thin
long-dashed curve indicates the existence of two retrieval states with
different free energies. Both stable retrieval states have the same free
energies at the thick long-dashed curves.   
In order to get an idea about the relevant $b$ and m-values for this
case, we note that for $\alpha = 0$
\begin{equation}
\begin{tabular}{rcccllclcl}
    &   & b &$<$&1/4 &:\hskip2cm & $m=$ & $6/5$ &          &    \\
 1/4&$<$& b &$<$&$3/10$&:\hskip2cm & $m=$ & $6/5$   &{\rm and}
                      & 1 \\
 $3/10$&$<$& b &$<$&3/4 &:\hskip2cm & $m=$ & 1   &{\rm and}
                  & $2/5$ \\
 3/4&$<$& b &  &     &:\hskip2cm & $m=$ & $2/5$ &          &
\end{tabular}
\label{eq:hr5}
\end{equation}
Again, these results are valid independent of the architecture (compare
\cite{bsvz94,brs94}).

Concerning the spin-glass states we have to solve the equations 
(\ref{q4mt0})-(\ref{q4ct0}) for $m=0$. It is straightforward to check that 
a spin-glass phase 
exists in the whole region of the phase diagram. By looking at the free
energy expression (\ref{feT0}) we find that it is always energetically  
unfavourable versus the retrieval state (only versus the one with the lower 
free energy in case there are two). Furthermore, the retrieval state is
unstable against replica symmetry breaking.

Comparing the phase diagram of the $Q=4$ model with that of 
the asymmetrically diluted model \cite{bsvz94} we see that similar to
the $Q=3$ case there is an extended $2$-Ising-like region here. Again
the overall shape is similar apart from the fact that it is tilted towards 
greater $b$-values (compare \cite{bsvz94} fig.~4).

\subsection{$Q=\infty$}

Finally we turn to the case $Q=\infty$. Considering again uniformly
distributed patterns between $-1$ and 1 (and hence $A=1/3$) the fixed-point
equations are still given by (\ref{eq:s6})-(\ref{eq:s8}) with (\ref{eq:s9})
replaced by
\begin{equation}
       \av{\sigma(z)} = \frac{\int_{-1}^1d\sigma \, \sigma \exp[
           \beta \sigma( \sum_\mu m_\mu \xi^\mu+\sqrt{\alpha r}\,z
           -\tilde{b} \sigma)]}{ \int_{-1}^1d s\,\exp[
    \beta s(\sum_\mu m_\mu \xi^\mu+\sqrt{\alpha r}\,z-\tilde{b} s)]}\,.
        \label{eq:101}
\end{equation}

For the retrieval state at zero temperature this leads to the explicit 
fixed-point equations presented in (\ref{minf})-(\ref{T0chi}) of the Appendix.
These equations are written down for $\tilde{b}>0$.
Again, for $\tilde{b}\leq0$ the equations can be further reduced 
by introducing the variable $x= m/\sqrt{2\alpha q}$:
\begin{equation}
     \sqrt{2\alpha} = \frac32\left[\frac{\erf(x)}{x}\left(1-\frac{1}{2x^2}
                   \right)+\frac{1}{\sqrt{\pi}\,x^2}\exp(-x^2)\right]\,,
     \label{eq:105}
\end{equation}
together with, in view of the definition of $\tilde{b}$, the following 
condition
\begin{equation}
       b \leq b_0 =\sqrt{\frac{\alpha}{8}}\,\frac{\erf(x)}{x}\,.
       \label{eq:106}
\end{equation}
Exactly as in the $Q=3$ and $4$ model, these equations tell us that the
retrieval state vanishes continuously at $\alpha_c=2/{\pi}$ for 
$b \,\leq\, b_0 \,=\,1/{\pi}$. In fact, one can analytically show that
the critical boundary of this 2-Ising like region is independent of $Q$. 
Indeed, in this region the general fixed-point equations
(\ref{eq:s6})-(\ref{eq:s9}) at $T=0$ 
together with the condition $\tilde{b} \leq 0$ lead to     
\begin{eqnarray}
     \sqrt{2\alpha} = \frac{1}{A x} \left\langle\kern-0.2em\left\langle 
	 \xi \erf(x \xi) \right\rangle\kern-0.2em\right\rangle 
	\label{like}    \\
      b \leq \sqrt{\frac{\alpha}{2\pi}} 
      \left\langle\kern-0.2em\left\langle 
          \exp(-x^2 \xi^2) \right\rangle\kern-0.2em\right\rangle,
	  \quad x \in [0, \infty]\,.
\end{eqnarray}   
Noting that the r.h.s. of eq.~(\ref{like}) is monotonically
decreasing for $x>0$, one immediately concludes that the phase 
boundary for this state is given by $x\rightarrow 0$ so that
$\alpha_c = 2/\pi$ and $b_0=1/\pi$, regardless of $Q$.    

\begin{figure}[t]
     \vspace{1.5cm} 
      \centering
       \includegraphics[height=0.5\textwidth,
                      angle=0]{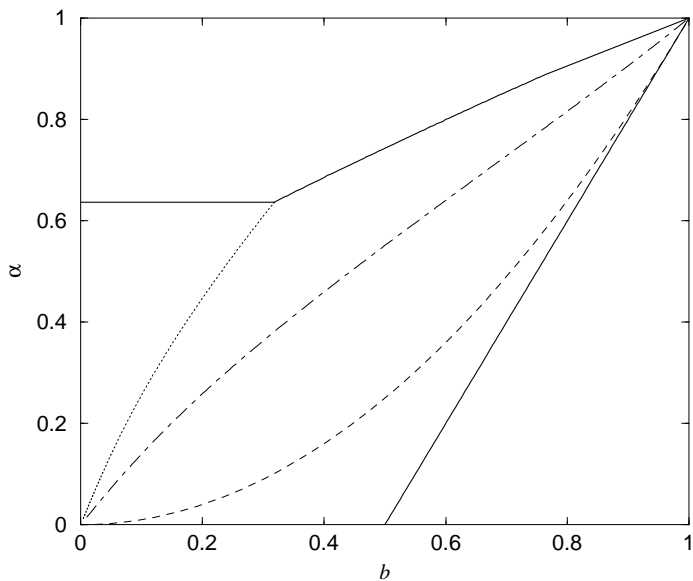}
       \caption{ $Q=\infty$ $\alpha-b$ phase diagram at $T=0$ for uniformly
       distributed patterns. The short-dashed line 
      indicates the continuous appearance of the spin-glass state.
      The meaning of the other lines is as in Figs.\ 1 and 2.}
       \label{fig:qifT0}
\end{figure}

For $\tilde{b}>0$ the region where the retrieval solution exists is found 
numerically by
solving the fixed-point equations (\ref{minf})-(\ref{T0chi}). The result
is shown in figure \ref{fig:qifT0} as the full line. The solution
appears continuously. 
Compared with all other architectures treated in the literature 
 --asymmetrically diluted, layered and fully connected--  where the storage
capacity is zero for $b \geq 1/2$, this is only the case here at finite 
loading $\alpha=0$. Because of the tilting of the phase diagram in comparison
with the one for the asymmetrically diluted model, as seen already for
$Q=3,4$, this is no longer the case for non-finite loading.
The retrieval state is the global minimum of the free energy in the
whole retrieval region.
   
For a discussion of the spin-glass phase one can follow a similar
argumentation as for the $Q=3$ model starting from the fixed-point 
equation analogous to eq.~(\ref{spinnew}).
One finds that the region of existence of the solution is bounded below
by $\alpha \geq b^2$, i.e. the dashed curve in figure \ref{fig:qifT0}. At 
the boundary $\chi=1/b$ and the spin-glass state appears continuously. 

In order to determine stability of the retrieval solution against replica 
symmetry breaking we
calculate the replicon eigenvalue  using eq.~(\ref{replinew}). The
result reads  
\begin{equation} 
       \lambda_R = 1 \,-\,
         \frac{\alpha}{4 \tilde{b}^2}
               \left\langle\kern-0.2em\left\langle
                 \int_{|m\xi +\sqrt{\alpha q} z |<2 \tilde{b}} {\cal D}z\, 
              \right\rangle\kern-0.2em\right\rangle_{\{\xi\}}  
	 \,\,
	     =  1 \,-\, \frac{\alpha \chi}{2 \tilde{b}}
	\label{replinf}
\end{equation} 
where we have used the fixed-point equation (\ref{T0chi}) for $\chi$.  
In contrast with the $Q=3$ model where the replicon eigenvalue is
always negative and hence breaking occurs in the whole retrieval region we
find that there is only partial breaking here. The AT  
boundary above which the replicon eigenvalue is positive and, hence,
breaking occurs, is shown as the dashed-dotted curve in 
figure \ref{fig:qifT0}.

\begin{figure}
       \centering   
       \subfigure[]{
             \includegraphics[height=0.45\textwidth,
                               angle=0]{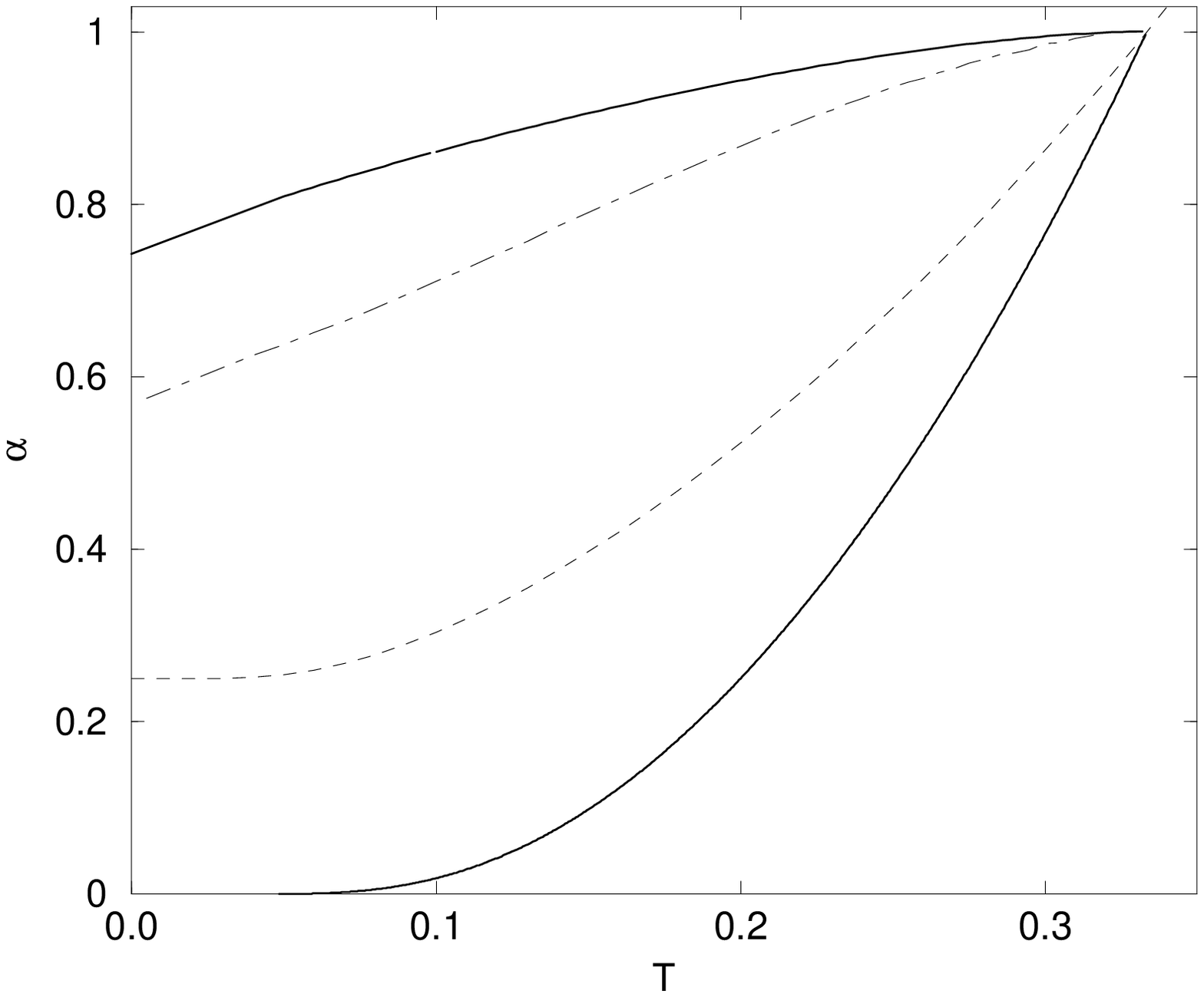}}
     	 \hspace{0.3cm}
	\subfigure[]{
             \includegraphics[height=0.45\textwidth,
                              angle=0]{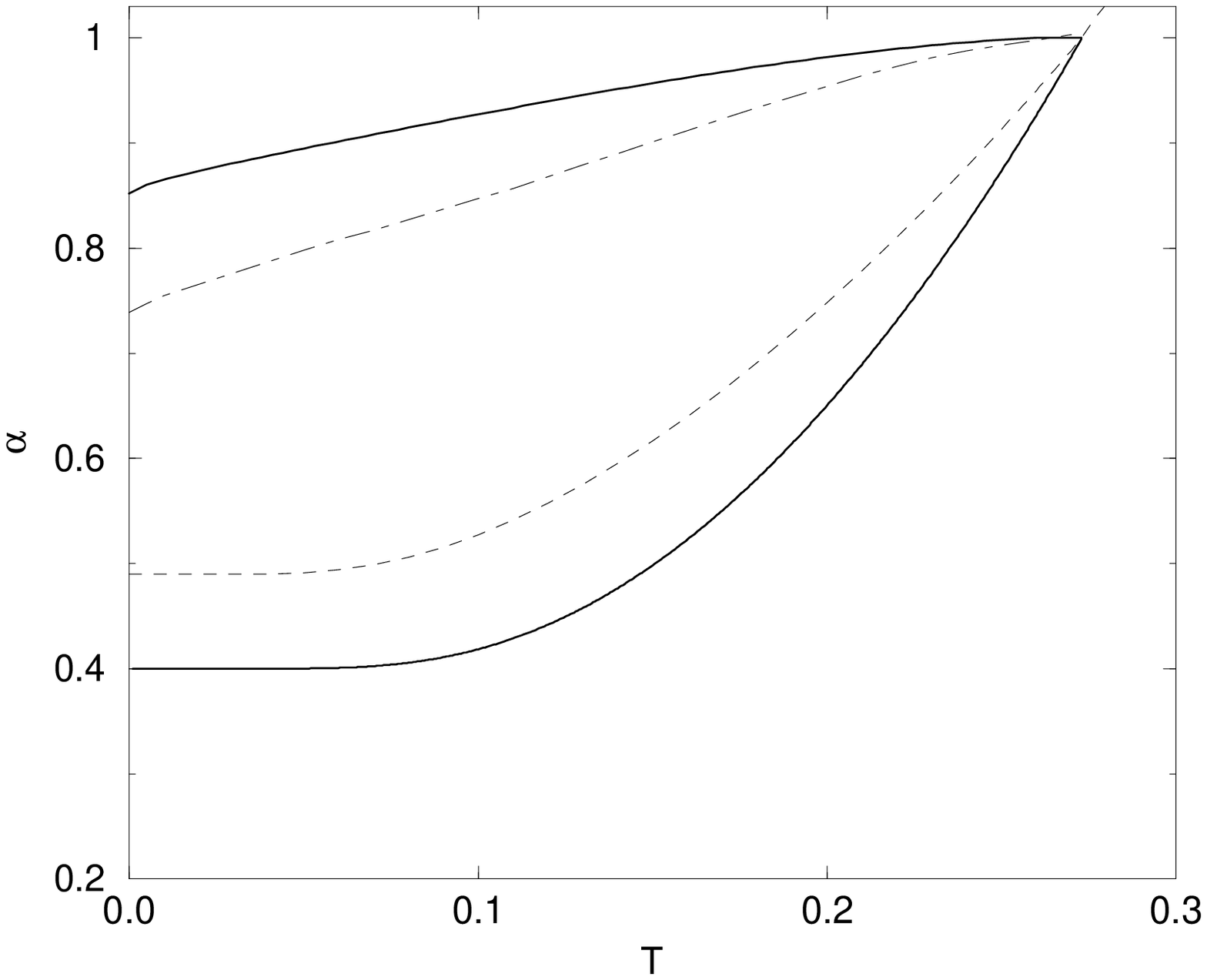}}     
	     \caption{ $Q=\infty$ $\alpha-T$ phase diagram for uniformly
	      distributed
       patterns and $(a)$: $b=0.5$ and $(b)$: $b=0.7$.The short-dashed line 
      indicates the continuous appearance of the spin-glass state.
       The meaning of the other lines is as in Figs.\ 1 and 2.} 
       \label{fig:qifTn}
\end{figure}

Finally, we study the $Q=\infty$ model at non-zero temperatures.
The retrieval phase is obtained by numerically solving the fixed-point
equations (\ref{eq:s6})-(\ref{eq:s9}). The result is presented as the
full line in figure \ref{fig:qifTn}. The retrieval state appears
continuously and
the lower branch of the retrieval boundary ends in zero for $b=0.5$. For
smaller $b$ it crosses the $T$-axis, e.g, at $T=1/3$ for $b=0$.

Next, we find the boundary of the spin-glass phase by expanding
the relevant equations (\ref{eq:s6})-(\ref{eq:s9}) for $m=0$ with respect
to $q$. We obtain
\begin{equation} 
     \alpha_{SG} \,=\,  \chi_0^{-2} 
\end{equation} 
with 
 \begin{equation}
   \chi_0 = \beta
   \frac{\int_{-1}^1 d\sigma \sigma^2 \exp(-\beta \tilde{b_0}\sigma^2) }
        {\int_{-1}^1 d\sigma  \exp(-\beta \tilde{b_0}\sigma^2 )}, \qquad
   \tilde{b_0} = b- \frac{1}{2 \chi_0}\,. 
\end{equation}
The result is the short-dashed line in figure \ref{fig:qifTn} above
which spin-glass solutions exist.  

Concerning the paramagnetic states the same reasoning can be followed as
for the $Q=3$ model after eq.~(\ref{para}) leading to one stable
solution, $\chi \sim 1/b$, with respect to longitudinal and transverse
fluctuations. However, this solution becomes
unstable against transverse fluctuations in the spin-glass region.

Comparing the relevant free energies we find that the retrieval states
are the global minima in the whole retrieval region. 
Finally, eq.~(\ref{replinew}) tells us that above the dashed-dotted
AT-line the replica symmetric retrieval solution is unstable. Again the
whole upper-branch of the retrieval boundary lies in this region.
Except for the relevant temperature scale these results show some 
qualitative similarities with the corresponding  $Q=3$ results.

\section{Concluding remarks}

We have considered both the thermodynamic properties and retrieval
properties of {\it symmetrically} diluted
$Q$-Ising networks. Fixed-point equations have been derived
for general temperature and arbitrary $Q$ in the replica-symmetric
mean-field approximation. For $Q=3,\,4,$ and $Q=\infty$ capacity-gain
parameter diagrams and capacity-temperature phase diagrams have been
discussed in detail. 

Concerning the capacity-gain parameter diagrams we find that the results
are essentially different for odd and even $Q$. Furthermore, there are
interesting similarities with the {\it asymmetric} extremely diluted
versions of the models. In fact, we find that the phase diagram here 
is tilted towards higher $b$-values because of the presence of 
an extended $2$-Ising-like region. The critical boundary of this region
is independent
of $Q$. Finally, the phase diagram for $Q=3$ is precisely the same as the 
one obtained using an exact dynamical approach after requiring that the 
evolution of the distribution of the local field becomes stationary.

Looking at the $\alpha-T$ phase diagrams we immediately notice the
overall qualitatively similar behaviour of the $Q=3$ and $Q=\infty$ model.
Furthermore the whole upper-branch of the retrieval boundary lies in the
replica symmetric unstable region, in contrast with the fully-connected
model.

\section*{Acknowledgments}

This work has been supported in part by the Research Fund of the
K.U. Leuven (grant OT/94/9). The authors are indebted to G.~Jongen
and G.~Massolo for stimulating discussions. One of us (D.B.) thanks the 
Fund for Scientific  Research - Flanders  (Belgium) for financial support.

\section*{Appendix}
\setcounter{equation}{0}
\renewcommand{\theequation}{A.\arabic{equation}}

In this appendix we write down explicitly the fixed-point equations
(\ref{eq:s6})-(\ref{eq:s9}) for $Q=3$, $Q=4$ and $Q=\infty$.
For a three-state network with patterns taking the values $0,\pm 1$ with
equal probability ,one obtains for a retrieval state
\begin{eqnarray}
      m &=& \int Dz\,V_\beta(m+\sqrt{\alpha q}\,z,\tilde{b})
      \label{eq:29}      \\
      q &=& \int Dz\,\left[\frac23 V_\beta^2(m+\sqrt{\alpha q}\,z,\tilde{b})
         + \frac13 V_\beta^2(\sqrt{\alpha q}\,z,\tilde{b}) \right] \\
    \chi &=& \frac{1}{\sqrt{\alpha q}} \int Dz \,z
            \left[\frac23V_\beta(m+\sqrt{\alpha q}\,z,\tilde{b})
            +\frac13 V_\beta(\sqrt{\alpha q}\,z,\tilde{b})\right]\,,
       \label{eq:31}
\end{eqnarray}
where
\begin{equation}
      V_\beta(x,y) \equiv
             \frac{\sinh(\beta x)}{\frac12\exp(\beta y)+\cosh(\beta x)}
        \label{eq:32}
\end{equation}
For zero temperature the Gaussian variable $z$ can be integrated out 
explicitly. For $\tilde{b}>0$ one arrives at
\begin{eqnarray}
 \fl    m &=& \frac12\biggl[ 
              \erf\biggl( \frac{m+\tilde{b}}{\sqrt{2\alpha q}} \biggr)
               +
              \erf\biggl( \frac{m-\tilde{b}}{\sqrt{2\alpha q}} \biggr)
           \biggr]
    \label{q3T0spe1}
            \\
 \fl    q &=& 1-\frac13\biggl[
              \erf\biggl( \frac{m+\tilde{b}}{\sqrt{2\alpha q}} \biggr)
              - \erf\biggl( \frac{m-\tilde{b}}{\sqrt{2\alpha q}} \biggr)
              + \erf\biggl( \frac{\tilde{b}}{\sqrt{2\alpha q}} \biggr)
           \biggr]
               \\
 \fl   \chi &=& \sqrt{\frac{2}{9\pi\alpha q}} \biggl[
                 \exp\biggl( -\frac{(m+\tilde{b})^2}{2\alpha q} \biggr)
                 + \exp\biggl( -\frac{(m-\tilde{b})^2}{2\alpha q} \biggr)
                 + \exp\biggl(-\frac{\tilde{b}^2}{2\alpha q} \biggr)
          \biggr]
   \,. \label{q3T0spe3}
\end{eqnarray}

For a four-state model in which the patterns
can take the value $\pm1/3, \pm1$ with equal probability  the fixed-point 
equations (\ref{eq:s6})-(\ref{eq:s9}) for a retrieval state at zero
temperature read
\begin{eqnarray}
\fl    m &=& \frac{3}{10} \biggl[ 
            \erf\biggl(\frac{3m+\tilde{b}}{3\sqrt{2\alpha q}} \biggr)
           +\erf\biggl(\frac{3m-\tilde{b}}{3\sqrt{2\alpha q}} \biggr)
           +\erf\biggl(\frac{m}{\sqrt{2\alpha q}} \biggr)
         \biggr]
             \nonumber \\
\fl      &&   +\frac{1}{10} \biggl[
            \erf\biggl(\frac{m+\tilde{b}}{3\sqrt{2\alpha q}} \biggr)
           +\erf\biggl( \frac{m-\tilde{b}}{3\sqrt{2\alpha q}} \biggr)
           +\erf\biggl( \frac{m}{3\sqrt{2\alpha q}} \biggr)
         \biggr]
           \label{q4mt0}  \\
\fl    q &=& 1 - \frac{2}{9}\biggl[ 
            \erf\biggl(\frac{3m+\tilde{b}}{3\sqrt{2\alpha q}} \biggr)
           -\erf\biggl(\frac{3m-\tilde{b}}{3\sqrt{2\alpha q}} \biggr)
           +\erf\biggl(\frac{m+\tilde{b}}{3\sqrt{2\alpha q}} \biggr)
           -\erf\biggl(\frac{m-\tilde{b}}{3\sqrt{2\alpha q}} \biggr)
           \biggr]
            \\
\fl     \chi &=& \frac{2}{3\sqrt{2\pi\alpha q}} \biggl[ 
            \exp\biggl(-\frac{(3m+\tilde{b})^2}{18\alpha q} \biggr)
           +\exp\biggl(-\frac{(3m-\tilde{b})^2}{18\alpha q} \biggr)
           +\exp\biggl(-\frac{m^2}{2\alpha q} \biggr)
          \nonumber \\
\fl         &&  +\exp\biggl(-\frac{(m+\tilde{b})^2}{18\alpha q} \biggr)
           +\exp\biggl(-\frac{(m-\tilde{b})^2}{18\alpha q} \biggr)
           +\exp\biggl(-\frac{m^2}{18\alpha q} \biggr)
           \biggr]
	   \label{q4ct0}
\end{eqnarray}
Again, the above expressions are valid only for $\tilde{b}\ge0$.

Finally, for $Q=\infty$ and uniformly distributed patterns
($A=1/{3}$)
a retrieval state is given by the solution of the fixed-point equations
(\ref{eq:s6})-(\ref{eq:s8}) and (\ref{eq:101}).
For the retrieval state at zero temperature the neuron expectation value
$\av{\sigma(z)}= g_{\tilde{b}}(m\xi+\sqrt{\alpha q}\,z)$, with the 
effective input-output
function $\tilde{g}$ given by (\ref{eq:r7}). This allows us to
perform explicitly the Gaussian average in the fixed-point equations 
resulting in
\begin{eqnarray} 
\fl   m &=&
      \frac{3}{2}\int_{-1}^{1} d\xi\,\xi\,
        \left[\left( 1+ \frac{m\xi}{2\tilde{b}} \right)\,
      {\rm Erf}\left(\frac{2\tilde{b}+m\xi}{\sqrt{2\alpha q}}\right)
      \,+\,
      \frac{1}{\tilde{b}} \sqrt{\frac{\alpha q}{2\pi}}
           \exp\left(-\frac{(2\tilde{b}+m\xi)^2}{2\alpha q}\right)\right]
     \label{minf} \\
\fl   q &=& 1\,+\,\frac{1}{2} \int_{-1}^{1} d\xi\,
  \left[ \left( \frac{\alpha q + (m\xi)^2}{(2\tilde{b})^2}-1 \right)\,
    {\rm Erf}\left(\frac{2\tilde{b}+m\xi}{\sqrt{2\alpha q}}\right) \right.
        \nonumber  \\ 
 \fl       && \left. + \frac{1}{\tilde{b}} 
      \sqrt{\frac{\alpha q}{2\pi}  }
       \left( \frac{m\xi}{2\tilde{b}} - 1 \right)
      \exp\left(-\frac{(2\tilde{b}+m\xi)^2}{2\alpha q}  \right) \right]  	
\label{T0q}
         \\  
\fl  \chi &=& \frac{1}{4 \tilde{b}}\, \int_{-1}^{1} d\xi\,
        {\rm Erf}\left(\frac{2\tilde{b}+m\xi}{\sqrt{2\alpha q}}\right)
     \label{T0chi}
\end{eqnarray}
for positive $\tilde{b}$. We remark that it is also
straightforward to perform the integration associated with the
random patterns but there is no need to write down the resulting 
expressions.

\newpage

\end{document}